\documentclass{iau}
\usepackage{graphicx}
\usepackage{natbib}

\title[chemistry \& GCs of MW dwarfs] %% give here short title %%
{Chemical Abundances and Globular Clusters of Milky Way Dwarf Galaxies}

\author[Tang et al.]   %% give here short author list %%
{Baitian Tang$^{1,2}$,
  Shihui Lin$^{1,2}$,  Ruoyun Huang$^{1,2}$, %Genghao Liu$^{1,2}$,
  Cheng Xu$^{1,2}$, %Yi Qiao$^{1,2}$, 
 Jos\'e G. Fern\'andez-Trincado$^{3,4}$,
 Doug Geisler$^{5,6}$, Chengyuan Li$^{1,2}$,
 Zhiqiang Yan$^{7,8}$%, Zhiyu Zhang$^{7,8}$, Guy Worthey$^{9}$, \and  Dante Minniti$^{10,11}$
 }

\affiliation{$^1$School of Physics and Astronomy, Sun Yat-sen University, Zhuhai 519082, People’s Republic of China  email: {\tt tangbt@mail.sysu.edu.cn} \\[\affilskip]
$^2$CSST Science Center for the Guangdong-Hong Kong-Macau Greater Bay Area, Zhuhai 519082, People’s Republic of China \\[\affilskip]
$^3${Universidad Cat\'olica del Norte, N\'ucleo UCN en Arqueolog\'ia Gal\'actica - Inst. de Astronom\'ia, Av. Angamos 0610, Antofagasta, Chile}\\[\affilskip]
$^4${Universidad Cat\'olica del Norte, Departamento de Ingenier\'ia de Sistemas y Computaci\'on, Av. Angamos 0610, Antofagasta, Chile}\\[\affilskip]
$^5${Departamento de Astronom\'{i}a, Casilla 160-C, Universidad de Concepci\'{o}n, Concepci\'{o}n, Chile}\\[\affilskip]
$^6${Departamento de Astronom\'ia, Facultad de Ciencias, Universidad de La Serena. Av. Ra\'ul Bitr\'an 1305, La Serena, Chile}\\[\affilskip]
$^7${School of Astronomy and Space Science, Nanjing University, Nanjing 210093, PR China}\\[\affilskip]
$^8${Key Laboratory of Modern Astronomy and Astrophysics (Nanjing University), Ministry of Education, Nanjing 210093, PR China}\\[\affilskip]
}

\pubyear{2025}
\volume{403}  %% insert here IAU Symposium No.
\pagerange{119--126}
\jname{The Hidden Beauty of the Galactic Outskirts}
\editors{**}
\begin{document}

\maketitle

\begin{abstract}
We present an overview of our ongoing GASTRONOMI project, which investigates the coevolution of the Milky Way (MW), its satellite dwarf galaxies, and their star clusters through chemo-dynamical analysis. We derive precise chemical abundances for stars in five classical dwarf galaxies, which reveal mass-dependent chemical evolution, particularly in $\alpha-$elements, such as [Si/Fe]. A distinct dichotomy in [Al/Fe] is found between metal-rich ($\mathrm{[Fe/H]} > -1.5$) stars formed in-situ in the MW and those originating in dwarf galaxies.

Star clusters act as sensitive environmental probes. The presence of multiple populations correlates with galactic evolution, and nitrogen-rich stars in Fornax are likely relics of disrupted globular clusters (GCs). We developed a chemical classification for Galactic GCs, isolating primordial populations by their low [Al/Fe]. This places in-situ and accreted GCs in distinct regions of the [Al/Fe]-[Fe/H] plane, providing a new tool to reconstruct the Galaxy’s accretion history.

%We demonstrate that star clusters are sensitive probes of their birth environment. The presence or absence of multiple populations (MPs) in clusters correlates with galactic evolution, and we identify nitrogen-rich field stars in Fornax as likely relics of disrupted globular clusters (GCs). Finally, we develop a novel, chemically based classification for Galactic GCs. By isolating the primordial population (lowest 1/3 in [Al/Fe]) within each cluster, we show that metal-rich in-situ and accreted GCs occupy distinct regions in the [Al/Fe]-[Fe/H] plane. This chemical classification generally agrees with dynamical (orbital) tagging methods but identifies potential outliers, offering a powerful new tool for reconstructing the accretion history of the Galaxy.
\keywords{Chemical Abundances, Globular Clusters, Dwarf Galaxies, Milky Way}
%% add here a maximum of 10 keywords, to be taken form the file <Keywords.txt>
\end{abstract}

\firstsection % if your document starts with a section,
              % remove some space above using this command.
\section{Introduction}

As the closest galaxy to us, the Milky Way (MW) provides an unparalleled testbed for studying galaxy formation and evolution. In this context, both fully accreted dwarf galaxies (e.g., Gaia-Sausage-Enceladus, or GSE, \citealt{2018MNRAS.478..611B,2018Natur.563...85H}) and those in the process of dissolving (e.g., Sagittarius, or Sgr, \citealt{2003ApJ...599.1082M}) are considered the building blocks of the present-day Galaxy. These accretion events contributed significantly to the Galactic mass and likely disturbed its dynamical structure \citep{2011Natur.477..301P}. More importantly, the ancient GSE merger may be linked to the formation of the Galactic thick disk \citep{2019NatAs...3..932G}, and the ``Splash'' population \citep{2020MNRAS.494.3880B}.
While member identification of these accreted dwarf galaxies may suffer from contamination, surviving dwarf galaxies (e.g, Sculptor, Fornax) provide clean samples for resolved stellar population studies. However, their large distances ($\sim$ a few 100 kpc) pose significant observational hurdles. Obtaining deep color-magnitude diagrams (CMD) that reach beyond their main sequence turn-offs or acquiring high-resolution spectra for precise chemical abundances has historically been difficult. In the early 2000s, only a handful of stars in each classical dwarf galaxy had been observed with high-resolution, high signal-to-noise (SNR)  spectra \citep{2003AJ....125..684S, 2005AJ....129.1428G}. Multi-object spectrograph technology has since revolutionized the field by rapidly expanding sample sizes. High SNR is now achieved either through large-aperture telescopes (e.g., VLT, \citealt{2019A&A...626A..15H}), or via long exposure on mid-sized telescopes (e.g., APOGEE, \citealt{Tang2023,  Xu2026}). The resulting precise chemical abundances have revealed the intricate star-formation histories of these dwarf galaxies.

Stars form within clusters, making these systems fundamental to galactic evolution.  The dense star clusters that survive to the present day serve as unique probes of the galactic conditions at their birth. Astronomers generally categorize star clusters into two groups: open clusters (OCs) and globular clusters (GCs). OCs are typically younger, less densely populated, and chemically homogeneous among member stars. In contrast, GCs are ancient, extremely dense, and exhibit significant internal variations in their chemical compositions. The observed chemical spreads (e.g., C, N, O, Na, Mg, Al, Fe-peak elements, and neutron-capture elements) in GCs are often referred as ``multiple populations'' (MPs).  GCs formed early and in great numbers within massive galaxies \citep{Harris2015}, with a substantial fraction dissolving over cosmic time. By leveraging their unique chemical features, e.g., N enrichment, recent studies have successfully identified a group of N-rich field stars as promising candidates of GC escaped stars \citep{FT2019, tang2019, tang2020, 2020ApJ...903L..17F}. Conversely, the GCs that survive intact can act as witnesses to galactic merger and accretion histories, as they are often brought into the central galaxy during these events. Determining their origins is therefore crucial for reconstructing the assembly history of their host galaxies.

\section{Chemical Abundances of MW Dwarf Galaxies}

\begin{figure}
% \vspace*{-2.0 cm}
\begin{center}
 \includegraphics[width=5.8in]{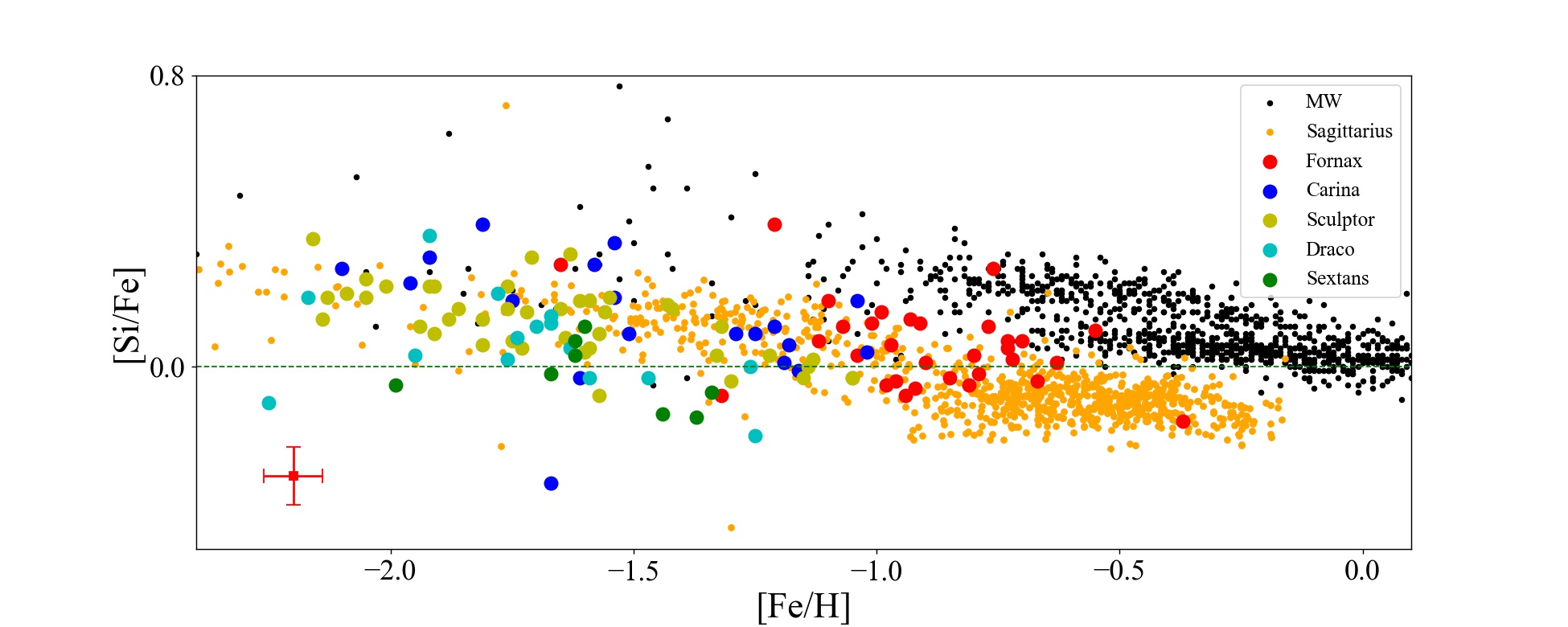} 
% \vspace*{-1.0 cm}
 \caption{[Si/Fe] vs. [Fe/H]. Scl, Fnx, Dra, Car and Sex stars from \citet{Tang2023} and \citet{Xu2026} are labeled as yellow-green, red, cyan, blue and green stars respectively. The error bars indicate the median uncertainties of available measurements. Black dots correspond to MW stars from the halo \citep{2000AJ....120.1841F,2004A&A...416.1117C,2005A&A...439..129B,2013ApJ...762...27Y,2014AJ....147..136R}, and MW stars from the disc \citep{2003MNRAS.340..304R,2006MNRAS.367.1329R,2014A&A...562A..71B}. Orange dots represent Sgr stars from \citet{2021ApJ...923..172H}. }
   \label{fig:alpha}
\end{center}
\end{figure}

We derived chemical abundances for individual stars in five classical dwarf galaxies --- Sculptor (Scl), Fornax (Fnx), Carina (Car), Draco (Dra), and Sextans (Sex) --- using high signal-to-noise (SNR$>70$), high-resolution ($R \sim 22 000$) near-infrared spectra from APOGEE \citep{Tang2023, Xu2026}. Our measurements include C, N, O, Mg, Al, Si, Ca, Ti, iron-peak elements, and Ce. These chemical abundances were measured using the Brussels Automatic Stellar Parameter (BACCHUS) code \citep{2016ascl.soft05004M}, which is built on the radiative transfer code, Turbospectrum \citep{1998A&A...330.1109A, 2012ascl.soft05004P}. We present Si and Al abundances as representative tracers of $\alpha$-elements and odd-Z elements, respectively. Figure \ref{fig:alpha} suggests that stars in low mass galaxies tend to be more metal-poor. Furthermore, the point where the fiducial [Si/Fe]-[Fe/H] line intersects [Si/Fe$]=0$ occurs at a lower [Fe/H] for galaxies of decreasing mass \citep{Xu2026}, highlighting the critical role of galaxy mass in shaping chemical evolution \citep{2009ARA&A..47..371T}. 

\begin{figure}
% \vspace*{-2.0 cm}
\begin{center}
 \includegraphics[width=4.0in]{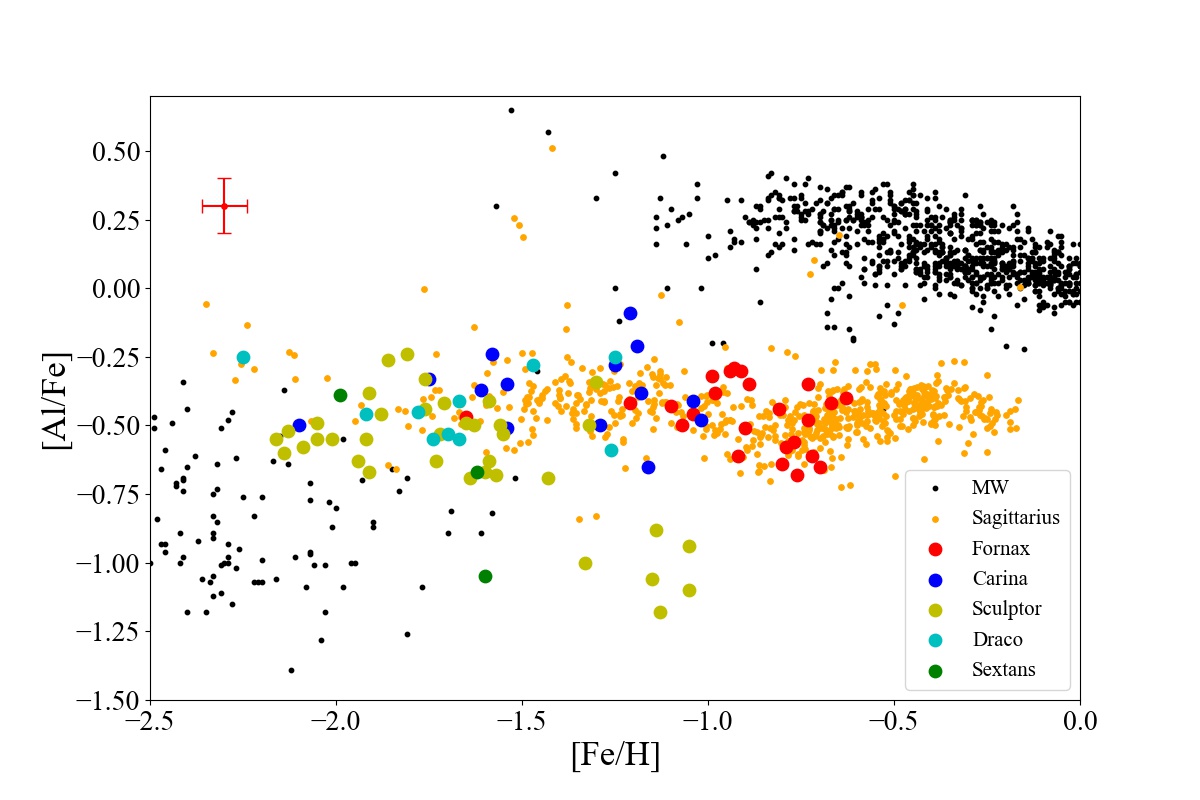} 
% \vspace*{-1.0 cm}
 \caption{[Al/Fe] vs. [Fe/H] relations. Symbols are the same as in Fig.~\ref{fig:alpha}.}
   \label{fig:Al}
\end{center}
\end{figure}

Figure \ref{fig:Al} suggests that stars in most dwarf galaxies generally show consistent [Al/Fe$]\sim-0.5$ across the metallicity range $-2.2<[$Fe/H$]<0$. Sgr presents interesting additional features, with clear changes in the slopes of its fiducial Si and Al trends around [Fe/H$]\sim-0.8$, likely indicative of secondary star formation episodes \citep{2021ApJ...923..172H}. A similar, though weaker, feature may be present in Fnx, though this is less distinct due to its smaller sample size. In contrast, MW stars show a different Al enrichment history. While metal-poor MW stars ([Fe/H$]<-1.5$) have [Al/Fe] ratios comparable to those in dwarf galaxies, more metal-rich MW stars ([Fe/H$]>-1.5$) are significantly enhanced, with [Al/Fe$]>-0.1$. This distinct chemical signature provides a powerful diagnostic for tracing the galactic origin of field stars \citep{das_ages_2020} and GCs (Section \ref{sect:Al}).

Radial chemical gradients within galaxies serve as important tracers for decoding spatially varying star formation histories. We investigated such gradients in Scl and Fnx by performing linear least-squares regression on [Fe/H] and several $\alpha$- and odd-Z element abundances ([O/Fe], [Mg/Fe], [Al/Fe], [Si/Fe], [Ca/Fe]). Our analysis confirmed the presence of a radial metallicity gradient in both galaxies. In Scl, we further detected statistically significant positive gradients for [Mg/Fe] and [Ca/Fe]. This spatial distribution aligns with photometric evidence from deep HST observations: \citet{betti_2019MNRAS.487.5862B} found that star formation persisted longer in the innermost region of Scl ($\sim 1.5$ Gyr) compared to the outskirts ($\sim 0.5$ Gyr), naturally leading to more pronounced chemical enrichment at smaller galactocentric distances. In contrast, we found no significant radial gradients for any of the measured abundances in Fnx. While this null result could be influenced by the relatively small sample size (32 stars) and measurement uncertainties, the intrinsic star formation history of Fnx likely plays a more fundamental role. Its proposed multiple epochs of star formation \citep{2006A&A...459..423B} and potential merger history \citep{2012ApJ...756L...2A} would have efficiently mixed the stellar populations, thereby erasing any strong radial chemical gradient.

To reveal Scl chemical evolution behind the measured abundances, we used the publicly available GCE code ``GalIMF'' \citep{2017A&A...607A.126Y,2019A&A...629A..93Y} to generate possible evolution tracks of element abundance ratios. The code adopts an empirically calibrated environment-dependent IMF variation law (the IGIMF theory). We adopt the yield tables in \citet{2010MNRAS.403.1413K}, \citet{2018ApJS..237...13L}, and \citet{1999ApJS..125..439I} for AGB, SN II, and SN Ia, respectively. More details of our GCE model are provided in \citet{Tang2023}. 

We assume a delayed-$\tau$ star formation history: 
\begin{equation}\label{eq:SFR}
    \bar{\psi}_{\rm 10~Myr}(t) = R \cdot t/\tau \cdot e^{-t/\tau},
\end{equation}
where $R$ and $\tau$ are characteristic SFR and star formation timescale constants. The best-fit results are $R=0.05~M_\odot/$yr and $\tau=150$~Myr. 
The resulting time-integrated galaxy-wide IMF has a top-light shape similar to the power-law IMF with a power index for massive stars of about $-2.7$. An SN Ia delay time of 100 Myr is preferred, in agreement with \citet{2022ApJ...925...66D}. The present-day stellar mass, $2.61\times 10^6~M_\odot$, and the mean stellar metallicity, ${\rm [Fe/H]}=-1.67$, of our best-fit model fit well with the literature values (\citealt{2011ApJ...727...78K}).

\section{Star Clusters in Galaxy Formation}
\label{sect:SC}

\begin{figure}
% \vspace*{-2.0 cm}
\begin{center}
 \includegraphics[width=5.6in]{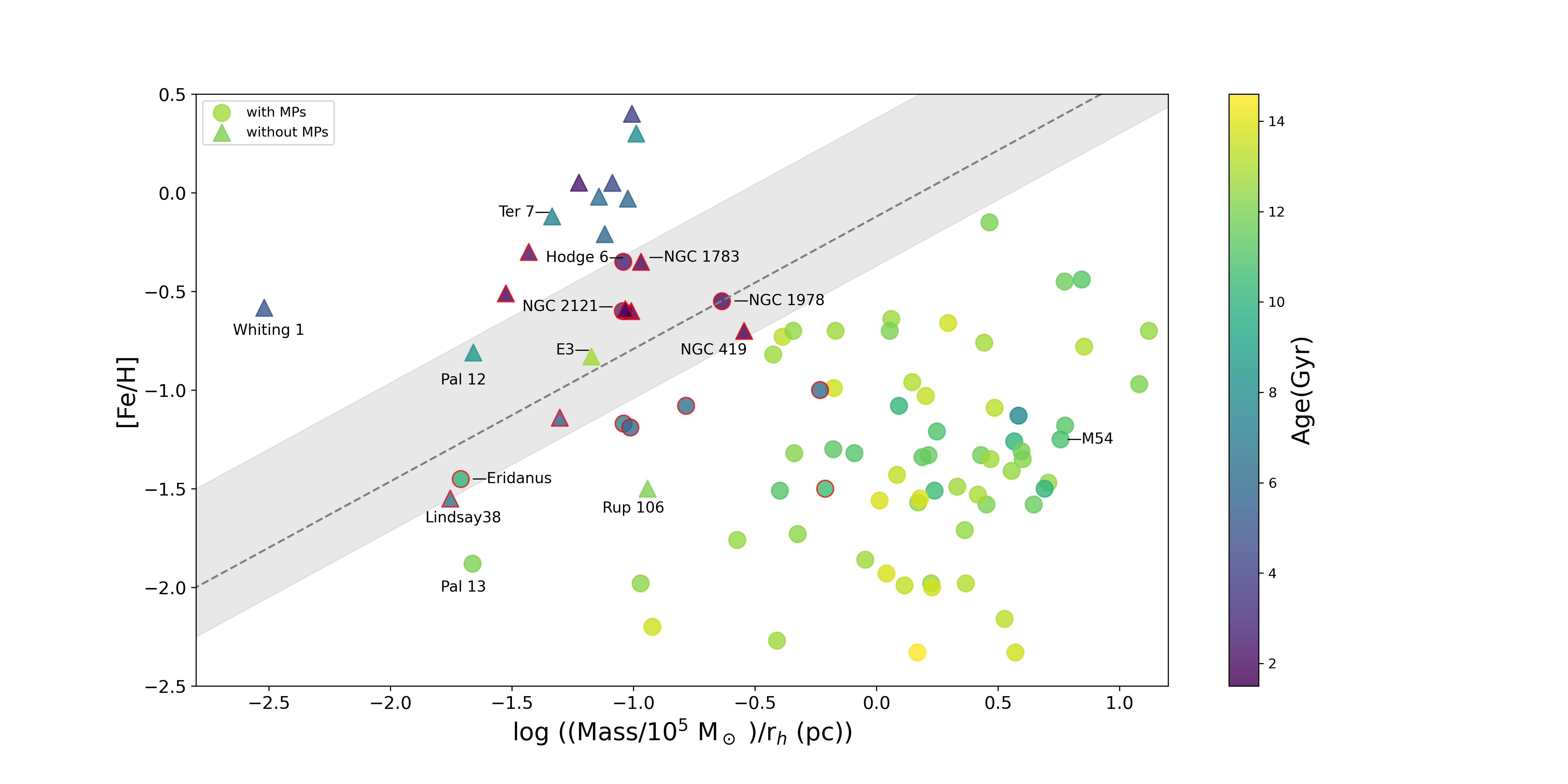} 
% \vspace*{-1.0 cm}
 \caption{ Star clusters older than 1.5 Gyr in the [Fe/H] vs. cluster compactness plane. Cluster compactness is defined as the ratio of the initial stellar mass M$_{\star}$ to the half-mass radius $r_h$, scaled by $10^5$ M$_{\odot}$ and in parsecs \citep{Krause2016}. Circles and triangles indicate clusters with and without MPs, respectively. Clusters with red edges are greater than 50 kpc from the galactic center. Their ages are indicated by their colors. The dashed line represents the proposed limit between galactic clusters with/without MPs. The shadow region represents the error after considering distant clusters. }
   \label{fig:C5}
\end{center}
\end{figure}

Star clusters, forming at specific times during a galaxy's evolution, are excellent probes of the galactic environment at their epoch of birth. For example, star clusters associated with the Sgr dwarf galaxy reveal multiple distinct star formation episodes in its history \citep{2010ApJ...718.1128L}. Recent analysis of the Sgr cluster Whiting 1 shows no sign of MP \citep{Huang2024}. When considered alongside other Sgr associated star clusters, a clear dichotomy emerges: clusters  formed early (e.g., M 54, Terzan 8) exhibit MPs, while those formed later (age$< 10$ Gyr, e.g., Whiting 1, Pal 12) do not. This pattern mirrors the dichotomy observed within the Galactic GC system. We propose that the underlying cause for this shift is the increasing metallicity of the galactic environment. As a galaxy evolves and becomes more metal-rich, the conditions necessary for the formation of MPs are likely no longer met. This interpretation is supported by our finding that stellar metallicity and cluster compactness are the key driving factors of the MP phenomenon  \citep[see Figure \ref{fig:C5},][]{Huang2024}.

Similarly, six star clusters have been identified in the Fnx dwarf galaxy. Among these, the earlier-formed GCs Fnx 1, Fnx 2, Fnx 3 \& Fnx 5 show chemical signatures of MPs, while the later-formed Fnx 4 \& Fnx 6 remain unexplored in this context. Fnx is particularly notable for its high GC specific frequency\footnote{The number of GCs per unit galaxy luminosity, i.e., $S_N= N_{GC} \times 10^{(M_V+15)}$.}, a feature that has attracted significant attention. Simulations suggest that the galaxy could have initially formed as many as $30-50$ GCs, depending on the models \citep[e.g.,][]{2023MNRAS.522.5638C}. The fact that only six survive today implies a substantial fraction have dissolved. In line with this picture, we identified 4 N-rich field stars within our samples of 32 Fnx stars ([N/Fe$]>0.65$, [C/Fe$]<0.15$, and [Ce/Fe$]<0.5$). Their elevated N abundances cannot be explained by extra-mixing, and instead closely resemble the distinctive chemical patterns of GC enrichment \citep{Xu2026}. These stars are thus strong candidates for being escapees from dissolved clusters.  Their metallicities ($-1.0<[$Fe/H$]<-0.8$)  do not match any of Fnx's surviving GCs, further supporting the interpretation that they originated in one or more now-disrupted GCs --- a scenario consistent with numerical simulations.

In contrast, our search yielded no N-rich field stars in low mass dwarf galaxies, i.e., Scl (out of 43 sample stars), Car (19), Dra (14), and Sex (8). Combined with the absence of any surviving GCs in these systems, this indicates a deficiency of the dense star forming environments required for GC formation. This finding is intriguingly consistent with independent evidence for a relative lack of massive stars in Scl, as inferred from its best-fit IMF, given that GCs are known to be efficient factories for high-mass star formation.

\begin{figure}
% \vspace*{-2.0 cm}
\begin{center}
 \includegraphics[width=4.2in]{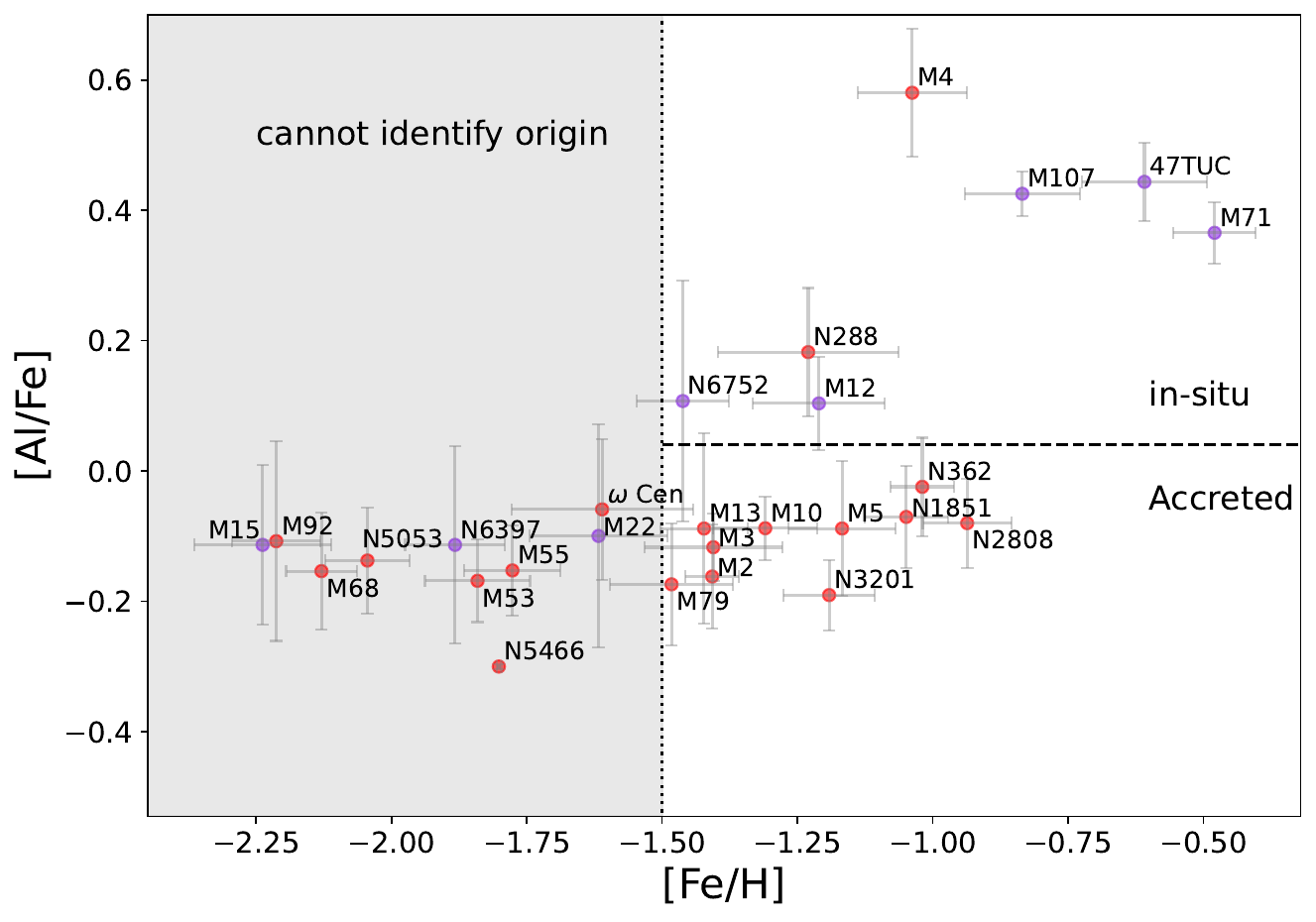} 
% \vspace*{-1.0 cm}
 \caption{ Mean [Al/Fe] versus mean [Fe/H] of primordial populations in Galactic GCs. Their associated standard deviations are shown as error bars. The primordial populations are defined as the lowest 1/3 in the [Al/Fe] distribution within each cluster. The black dotted line indicates the metallicity of [Fe/H$]=-1.5$. GCs with [Fe/H$]<-1.5$ are not suitable for our chemically based classification (grey region). The black dashed line separates chemically classified in-situ and accreted clusters. In comparison, MKH classification is also shown (in-situ: purple, accreted: red). }
   \label{fig:GCAl}
\end{center}
\end{figure}

\section{MW GC Classification Based on Their Mg-Al Abundances}
\label{sect:Al}

Galaxy mergers brought together GCs of different origins, assembling the Galactic GC population we observe today. Identifying their origins is crucial for constraining merger history and chemical evolution of the Galaxy. Thanks to the precise proper motions of GCs provided by the Gaia mission, clustering in their orbital parameter space links present-day Galactic GCs to their possible progenitor galaxies \citep[][, hereafter MKH]{Massari2019}. However, the assumption of invariant orbital parameters becomes tenuous when considering the Galaxy's complex merger history and long-term dynamical evolution. Simulations that include realistic ISM prescriptions also suggest that dynamical criteria alone may be problematic \citep[e.g.,][]{Pagnini2023}.

Unlike dynamical properties, chemical abundances are generally robust to galaxy mergers. Yet, chemically tagging GCs to their progenitors is complicated by the presence of MPs within GCs. A key question is whether it is possible to separate the chemical imprint of the host galaxy from that of internal GC enrichment. We hypothesize that for each GC, the lower envelope of its Al abundance distribution reflects the chemistry of its primordial population, which should retain the signature of its progenitor galaxy. 
Figure \ref{fig:Al} suggests that the lower bounds of Al distributions differ between  in-situ and accreted metal-rich ([Fe/H$]>-1.5$) GCs. To represent GC primordial populations while balancing subsample size, we selected the lowest 1/3 populations in the [Al/Fe] distribution for each GC. Using the homogeneous GC chemical abundances from \citet{Meszaros2020}, Figure \ref{fig:GCAl} confirms our initial speculation: metal-rich GCs ([Fe/H$]>-1.5$) from different galaxy origins show clear chemical dichotomy \citep{LinSH2025}. The primordial populations of in-situ GCs exhibit significantly higher [Al/Fe], consistent with the trend seen in Figure \ref{fig:Al}. To validate our classification, we compare it with the dynamically driven MKH scheme. The agreement is generally good, with two notable exceptions: NGC 288 and M4. Although classified as accreted by MKH, their primordial populations exhibit elevated [Al/Fe] abundances comparable to in-situ GCs. Interestingly, \citet{2025A&A...704A.256C} also proposed an in-situ origin for NGC 288, suggesting it was later dynamically heating by the GSE merger.

This chemical dichotomy stems from distinct nucleosynthetic pathways. As one of the odd-Z elements, Al is enhanced by the surplus of neutrons in $^{22}$Ne, and $^{22}$Ne is transformed from $^{14}$N by the CNO cycle during He burning. Consequently, Al yield correlates with the abundances of CNO elements, making it metallicity-dependent \citep[e.g., Fig. 5 of][]{K06}.  On the other hand, due to slow chemical enrichment in GSE-like galaxies, Type Ia supernova (SNe Ia) --- which produce substantial amounts of Fe --- begin to significantly influence galactic chemical evolution at comparatively low metallicities ([Fe/H$] \sim -1.5$). In contrast, SNe Ia began to significantly affect MW chemical evolution around [Fe/H$]\sim -0.8$. Thus, the observed dichotomy in [Al/Fe] among metal-rich ([Fe/H$]>-1.5$) GCs reflects distinct chemical evolutionary pathways between in-situ and accreted systems.

\section{Conclusion}

In this proceeding for IAU symposium, we present the working progress of our ongoing research initiative, ``Scrutinizing {\bf GA}laxy-{\bf ST}a{\bf R} cluster coevoluti{\bf ON} with chem{\bf O}dyna{\bf MI}cs \\({\bf \mbox{GASTRONOMI}})'', which leverages multi-wavelength photometric and spectroscopic data to unravel the coevolutionary relationships between the MW, its satellite dwarf galaxies, and their stellar clusters. Chemical abundances and their radial gradients serve as key diagnostics for decoding the star-formation histories and chemical evolution of dwarf galaxies. Star clusters, forming during galaxy assembly, preserve a record of the galactic environment at their birth. The observed transition from MP to no MP in star clusters of the Sgr and the MW indicates that cluster properties are affected by their host galactic environment, with low metallicity, compact clusters being especially conducive to MP formation. Furthermore, the detection of N-rich field stars in the Fnx dwarf galaxy may indicate the disruption of one or more GCs. Finally, we highlight our newly developed, chemically driven classification of GCs, which offers a promising tool for Galactic archaeology and the reconstruction of accretion history.

\begin{acknowledgements}
B.T., S.L., R.H. gratefully acknowledge support from the National Natural Science Foundation of China through grants NOs. 12473035 and 12233013, China Manned Space Project under grant NO. CMS-CSST-2025-A13 and CMS-CSST-2021-A08, the Fundamental Research Funds for the Central Universities, Sun Yat-sen University (24qnpy121). Deepseek is used for language improvements.
\end{acknowledgements}

%{\underline{\it Isotopic structures and nucleosynthesis}}. 

%\begin{discussion}

%\discuss{van der Hucht}{See the VIIth WR catalogue (van der Hucht 2001): of the listed 227
%Galactic WR stars, only 53 are in open culsters and OB associations, or believed to be. The other 184 are supposedly field stars.}
%\end{discussion}

\bibliographystyle{iaulike}
\bibliography{MW_satellite}

\begin{thebibliography}{}

\bibitem[{Alvarez} and {Plez}, 1998]{1998A&A...330.1109A}
{Alvarez}, R. \& {Plez}, B. 1998, {Near-infrared narrow-band photometry of
  M-giant and Mira stars: models meet observations}.
\newblock {\em \aap}, 330, 1109--1119.

\bibitem[{Amorisco} and {Evans}, 2012]{2012ApJ...756L...2A}
{Amorisco}, N.~C. \& {Evans}, N.~W. 2012, {A Troublesome Past: Chemodynamics of
  the Fornax Dwarf Spheroidal}.
\newblock {\em \apjl}, 756(1), L2.

\bibitem[{Barklem} et~al., 2005]{2005A&A...439..129B}
{Barklem}, P.~S., {Christlieb}, N., {Beers}, T.~C., {Hill}, V., {Bessell},
  M.~S., {Holmberg}, J., {Marsteller}, B., {Rossi}, S., {Zickgraf}, F.~J., \&
  {Reimers}, D. 2005, {The Hamburg/ESO R-process enhanced star survey (HERES).
  II. Spectroscopic analysis of the survey sample}.
\newblock {\em \aap}, 439(1), 129--151.

\bibitem[{Battaglia} et~al., 2006]{2006A&A...459..423B}
{Battaglia}, G., {Tolstoy}, E., {Helmi}, A., {Irwin}, M.~J., {Letarte}, B.,
  {Jablonka}, P., {Hill}, V., {Venn}, K.~A., {Shetrone}, M.~D., {Arimoto}, N.,
  {Primas}, F., {Kaufer}, A., {Francois}, P., {Szeifert}, T., {Abel}, T., \&
  {Sadakane}, K. 2006, {The DART imaging and CaT survey of the Fornax dwarf
  spheroidal galaxy}.
\newblock {\em \aap}, 459(2), 423--440.

\bibitem[{Belokurov} et~al., 2018]{2018MNRAS.478..611B}
{Belokurov}, V., {Erkal}, D., {Evans}, N.~W., {Koposov}, S.~E., \& {Deason},
  A.~J. 2018, {Co-formation of the disc and the stellar halo}.
\newblock {\em \mnras}, 478(1), 611--619.

\bibitem[{Belokurov} et~al., 2020]{2020MNRAS.494.3880B}
{Belokurov}, V., {Sanders}, J.~L., {Fattahi}, A., {Smith}, M.~C., {Deason},
  A.~J., {Evans}, N.~W., \& {Grand}, R. J.~J. 2020, {The biggest splash}.
\newblock {\em \mnras}, 494(3), 3880--3898.

\bibitem[{Bensby} et~al., 2014]{2014A&A...562A..71B}
{Bensby}, T., {Feltzing}, S., \& {Oey}, M.~S. 2014, {Exploring the Milky Way
  stellar disk. A detailed elemental abundance study of 714 F and G dwarf stars
  in the solar neighbourhood}.
\newblock {\em \aap}, 562, A71.

\bibitem[{Bettinelli} et~al., 2019]{betti_2019MNRAS.487.5862B}
{Bettinelli}, M., {Hidalgo}, S.~L., {Cassisi}, S., {Aparicio}, A., {Piotto},
  G., {Valdes}, F., \& {Walker}, A.~R. 2019, {The star formation history of the
  Sculptor dwarf spheroidal galaxy}.
\newblock {\em \mnras}, 487(4), 5862--5873.

\bibitem[{Cayrel} et~al., 2004]{2004A&A...416.1117C}
{Cayrel}, R., {Depagne}, E., {Spite}, M., {Hill}, V., {Spite}, F.,
  {Fran{\c{c}}ois}, P., {Plez}, B., {Beers}, T., {Primas}, F., {Andersen}, J.,
  {Barbuy}, B., {Bonifacio}, P., {Molaro}, P., \& {Nordstr{\"o}m}, B. 2004,
  {First stars V - Abundance patterns from C to Zn and supernova yields in the
  early Galaxy}.
\newblock {\em \aap}, 416, 1117--1138.

\bibitem[{Ceccarelli} et~al., 2025]{2025A&A...704A.256C}
{Ceccarelli}, E., {Massari}, D., {Aguado-Agelet}, F., {Mucciarelli}, A.,
  {Cassisi}, S., {Monelli}, M., {Pancino}, E., {Salaris}, M., \& {Saracino}, S.
  2025, {Cluster Ages to Reconstruct the Milky Way Assembly (CARMA): III. NGC
  288 as the first Splashed globular cluster}.
\newblock {\em \aap}, 704, A256.

\bibitem[{Chen} and {Gnedin}, 2023]{2023MNRAS.522.5638C}
{Chen}, Y. \& {Gnedin}, O.~Y. 2023, {Formation of globular clusters in dwarf
  galaxies of the Local Group}.
\newblock {\em \mnras}, 522(4), 5638--5653.

\bibitem[Das et~al., 2020]{das_ages_2020}
Das, P., Hawkins, K., \& Jofré, P. 2020, Ages and kinematics of chemically
  selected, accreted {Milky} {Way} halo stars.
\newblock {\em MNRAS}, 493(4), 5195--5207.

\bibitem[{de los Reyes} et~al., 2022]{2022ApJ...925...66D}
{de los Reyes}, M. A.~C., {Kirby}, E.~N., {Ji}, A.~P., \& {Nu{\~n}ez}, E.~H.
  2022, {Simultaneous Constraints on the Star Formation History and
  Nucleosynthesis of Sculptor dSph}.
\newblock {\em \apj}, 925(1), 66.

\bibitem[{Fern{\'a}ndez-Trincado} et~al., 2020]{2020ApJ...903L..17F}
{Fern{\'a}ndez-Trincado}, J.~G., {Beers}, T.~C., {Minniti}, D., {Carigi}, L.,
  {Barbuy}, B., {Placco}, V.~M., {Moni Bidin}, C., {Villanova}, S.,
  {Roman-Lopes}, A., \& {Nitschelm}, C. 2020, {Discovery of a Large Population
  of Nitrogen-enhanced Stars in the Magellanic Clouds}.
\newblock {\em \apjl}, 903(1), L17.

\bibitem[{Fern{\'a}ndez-Trincado} et~al., 2019]{FT2019}
{Fern{\'a}ndez-Trincado}, J.~G., {Beers}, T.~C., {Tang}, B., {Moreno}, E.,
  {P{\'e}rez-Villegas}, A., \& {Ortigoza-Urdaneta}, M. 2019, {Chemodynamics of
  newly identified giants with a globular cluster like abundance patterns in
  the bulge, disc, and halo of the Milky Way}.
\newblock {\em \mnras}, 488(2), 2864--2880.

\bibitem[{Fulbright}, 2000]{2000AJ....120.1841F}
{Fulbright}, J.~P. 2000, {Abundances and Kinematics of Field Halo and Disk
  Stars. I. Observational Data and Abundance Analysis}.
\newblock {\em \aj}, 120(4), 1841--1852.

\bibitem[{Gallart} et~al., 2019]{2019NatAs...3..932G}
{Gallart}, C., {Bernard}, E.~J., {Brook}, C.~B., {Ruiz-Lara}, T., {Cassisi},
  S., {Hill}, V., \& {Monelli}, M. 2019, {Uncovering the birth of the Milky Way
  through accurate stellar ages with Gaia}.
\newblock {\em Nature Astronomy}, 3, 932--939.

\bibitem[{Geisler} et~al., 2005]{2005AJ....129.1428G}
{Geisler}, D., {Smith}, V.~V., {Wallerstein}, G., {Gonzalez}, G., \&
  {Charbonnel}, C. 2005, {``Sculptor-ing'' the Galaxy? The Chemical
  Compositions of Red Giants in the Sculptor Dwarf Spheroidal Galaxy}.
\newblock {\em \aj}, 129(3), 1428--1442.

\bibitem[{Harris} et~al., 2015]{Harris2015}
{Harris}, W.~E., {Harris}, G.~L., \& {Hudson}, M.~J. 2015, {Dark Matter Halos
  in Galaxies and Globular Cluster Populations. II. Metallicity and
  Morphology}.
\newblock {\em \apj}, 806(1), 36.

\bibitem[{Hasselquist} et~al., 2021]{2021ApJ...923..172H}
{Hasselquist}, S., {Hayes}, C.~R., {Lian}, J., {Weinberg}, D.~H., {Zasowski},
  G., {Horta}, D., {Beaton}, R., {Feuillet}, D.~K., {Garro}, E.~R., {Gallart},
  C., {Smith}, V.~V., {Holtzman}, J.~A., {Minniti}, D., {Lacerna}, I.,
  {Shetrone}, M., {J{\"o}nsson}, H., {Cioni}, M.-R.~L., {Fillingham}, S.~P.,
  {Cunha}, K., {O'Connell}, R., {Fern{\'a}ndez-Trincado}, J.~G., {Mu{\~n}oz},
  R.~R., {Schiavon}, R., {Almeida}, A., {Anguiano}, B., {Beers}, T.~C.,
  {Bizyaev}, D., {Brownstein}, J.~R., {Cohen}, R.~E., {Frinchaboy}, P.,
  {Garc{\'\i}a-Hern{\'a}ndez}, D.~A., {Geisler}, D., {Lane}, R.~R., {Majewski},
  S.~R., {Nidever}, D.~L., {Nitschelm}, C., {Povick}, J., {Price-Whelan}, A.,
  {Roman-Lopes}, A., {Rosado}, M., {Sobeck}, J., {Stringfellow}, G.,
  {Valenzuela}, O., {Villanova}, S., \& {Vincenzo}, F. 2021, {APOGEE Chemical
  Abundance Patterns of the Massive Milky Way Satellites}.
\newblock {\em \apj}, 923(2), 172.

\bibitem[{Helmi} et~al., 2018]{2018Natur.563...85H}
{Helmi}, A., {Babusiaux}, C., {Koppelman}, H.~H., {Massari}, D., {Veljanoski},
  J., \& {Brown}, A. G.~A. 2018, {The merger that led to the formation of the
  Milky Way's inner stellar halo and thick disk}.
\newblock {\em \nat}, 563(7729), 85--88.

\bibitem[{Hill} et~al., 2019]{2019A&A...626A..15H}
{Hill}, V., {Sk{\'u}lad{\'o}ttir}, {\'A}., {Tolstoy}, E., {Venn}, K.~A.,
  {Shetrone}, M.~D., {Jablonka}, P., {Primas}, F., {Battaglia}, G., {de Boer},
  T.~J.~L., {Fran{\c{c}}ois}, P., {Helmi}, A., {Kaufer}, A., {Letarte}, B.,
  {Starkenburg}, E., \& {Spite}, M. 2019, {VLT/FLAMES high-resolution chemical
  abundances in Sculptor: a textbook dwarf spheroidal galaxy}.
\newblock {\em \aap}, 626, A15.

\bibitem[{Huang} et~al., 2024]{Huang2024}
{Huang}, R., {Tang}, B., {Li}, C., {Geisler}, D., {Mateo}, M., {Song}, Y.-Y.,
  {Baumgardt}, H., {Carballo-Bello}, J.~A., {Wang}, Y., {Nie}, J., {Dias}, B.,
  \& {Fern{\'a}ndez-Trincado}, J.~G. 2024, {Driving factors behind multiple
  populations}.
\newblock {\em Science China Physics, Mechanics, and Astronomy}, 67(5), 259513.

\bibitem[{Iwamoto} et~al., 1999]{1999ApJS..125..439I}
{Iwamoto}, K., {Brachwitz}, F., {Nomoto}, K., {Kishimoto}, N., {Umeda}, H.,
  {Hix}, W.~R., \& {Thielemann}, F.-K. 1999, {Nucleosynthesis in Chandrasekhar
  Mass Models for Type IA Supernovae and Constraints on Progenitor Systems and
  Burning-Front Propagation}.
\newblock {\em \apjs}, 125(2), 439--462.

\bibitem[{Karakas}, 2010]{2010MNRAS.403.1413K}
{Karakas}, A.~I. 2010, {Updated stellar yields from asymptotic giant branch
  models}.
\newblock {\em \mnras}, 403(3), 1413--1425.

\bibitem[{Kirby} et~al., 2011]{2011ApJ...727...78K}
{Kirby}, E.~N., {Lanfranchi}, G.~A., {Simon}, J.~D., {Cohen}, J.~G., \&
  {Guhathakurta}, P. 2011, {Multi-element Abundance Measurements from
  Medium-resolution Spectra. III. Metallicity Distributions of Milky Way Dwarf
  Satellite Galaxies}.
\newblock {\em \apj}, 727(2), 78.

\bibitem[{Kobayashi} et~al., 2006]{K06}
{Kobayashi}, C., {Umeda}, H., {Nomoto}, K., {Tominaga}, N., \& {Ohkubo}, T.
  2006, {Galactic Chemical Evolution: Carbon through Zinc}.
\newblock {\em \apj}, 653(2), 1145--1171.

\bibitem[{Krause} et~al., 2016]{Krause2016}
{Krause}, M. G.~H., {Charbonnel}, C., {Bastian}, N., \& {Diehl}, R. 2016, {Gas
  expulsion in massive star clusters?. Constraints from observations of young
  and gas-free objects}.
\newblock {\em \aap}, 587, A53.

\bibitem[{Law} and {Majewski}, 2010]{2010ApJ...718.1128L}
{Law}, D.~R. \& {Majewski}, S.~R. 2010, {Assessing the Milky Way Satellites
  Associated with the Sagittarius Dwarf Spheroidal Galaxy}.
\newblock {\em \apj}, 718(2), 1128--1150.

\bibitem[{Limongi} and {Chieffi}, 2018]{2018ApJS..237...13L}
{Limongi}, M. \& {Chieffi}, A. 2018, {Presupernova Evolution and Explosive
  Nucleosynthesis of Rotating Massive Stars in the Metallicity Range -3
  {\ensuremath{\leq}} [Fe/H] {\ensuremath{\leq}} 0}.
\newblock {\em \apjs}, 237(1), 13.

\bibitem[{Lin} et~al., 2025]{LinSH2025}
{Lin}, S., {Tang}, B., {Liu}, G., {Fern{\'a}ndez-Trincado}, J.~G., {Geisler},
  D., {Worthey}, G., \& {Minniti}, D. 2025, {Revealing the Origins of Galactic
  Globular Clusters via their Mg{\textendash}Al Abundances}.
\newblock {\em \apjl}, 989(2), L37.

\bibitem[{Majewski} et~al., 2003]{2003ApJ...599.1082M}
{Majewski}, S.~R., {Skrutskie}, M.~F., {Weinberg}, M.~D., \& {Ostheimer}, J.~C.
  2003, {A Two Micron All Sky Survey View of the Sagittarius Dwarf Galaxy. I.
  Morphology of the Sagittarius Core and Tidal Arms}.
\newblock {\em \apj}, 599(2), 1082--1115.

\bibitem[{Massari} et~al., 2019]{Massari2019}
{Massari}, D., {Koppelman}, H.~H., \& {Helmi}, A. 2019, {Origin of the system
  of globular clusters in the Milky Way}.
\newblock {\em \aap}, 630, L4.

\bibitem[{Masseron} et~al., 2016]{2016ascl.soft05004M}
{Masseron}, T., {Merle}, T., \& {Hawkins}, K. 2016,.
\newblock {BACCHUS: Brussels Automatic Code for Characterizing High accUracy
  Spectra}.
\newblock Astrophysics Source Code Library, record ascl:1605.004.

\bibitem[{M{\'e}sz{\'a}ros} et~al., 2020]{Meszaros2020}
{M{\'e}sz{\'a}ros}, S., {Masseron}, T., {Garc{\'\i}a-Hern{\'a}ndez}, D.~A.,
  {Allende Prieto}, C., {Beers}, T.~C., {Bizyaev}, D., {Chojnowski}, D.,
  {Cohen}, R.~E., {Cunha}, K., {Dell'Agli}, F., {Ebelke}, G.,
  {Fern{\'a}ndez-Trincado}, J.~G., {Frinchaboy}, P., {Geisler}, D.,
  {Hasselquist}, S., {Hearty}, F., {Holtzman}, J., {Johnson}, J., {Lane},
  R.~R., {Lacerna}, I., {Longa-Pe{\~n}a}, P., {Majewski}, S.~R., {Martell},
  S.~L., {Minniti}, D., {Nataf}, D., {Nidever}, D.~L., {Pan}, K., {Schiavon},
  R.~P., {Shetrone}, M., {Smith}, V.~V., {Sobeck}, J.~S., {Stringfellow},
  G.~S., {Szigeti}, L., {Tang}, B., {Wilson}, J.~C., \& {Zamora}, O. 2020,
  {Homogeneous analysis of globular clusters from the APOGEE survey with the
  BACCHUS code - II. The Southern clusters and overview}.
\newblock {\em \mnras}, 492(2), 1641--1670.

\bibitem[{Pagnini} et~al., 2023]{Pagnini2023}
{Pagnini}, G., {Di Matteo}, P., {Khoperskov}, S., {Mastrobuono-Battisti}, A.,
  {Haywood}, M., {Renaud}, F., \& {Combes}, F. 2023, {The distribution of
  globular clusters in kinematic spaces does not trace the accretion history of
  the host galaxy}.
\newblock {\em \aap}, 673, A86.

\bibitem[{Plez}, 2012]{2012ascl.soft05004P}
{Plez}, B. 2012,.
\newblock {Turbospectrum: Code for spectral synthesis}.
\newblock Astrophysics Source Code Library, record ascl:1205.004.

\bibitem[{Purcell} et~al., 2011]{2011Natur.477..301P}
{Purcell}, C.~W., {Bullock}, J.~S., {Tollerud}, E.~J., {Rocha}, M., \&
  {Chakrabarti}, S. 2011, {The Sagittarius impact as an architect of spirality
  and outer rings in the Milky Way}.
\newblock {\em \nat}, 477(7364), 301--303.

\bibitem[{Reddy} et~al., 2006]{2006MNRAS.367.1329R}
{Reddy}, B.~E., {Lambert}, D.~L., \& {Allende Prieto}, C. 2006, {Elemental
  abundance survey of the Galactic thick disc}.
\newblock {\em \mnras}, 367(4), 1329--1366.

\bibitem[{Reddy} et~al., 2003]{2003MNRAS.340..304R}
{Reddy}, B.~E., {Tomkin}, J., {Lambert}, D.~L., \& {Allende Prieto}, C. 2003,
  {The chemical compositions of Galactic disc F and G dwarfs}.
\newblock {\em \mnras}, 340(1), 304--340.

\bibitem[{Roederer} et~al., 2014]{2014AJ....147..136R}
{Roederer}, I.~U., {Preston}, G.~W., {Thompson}, I.~B., {Shectman}, S.~A.,
  {Sneden}, C., {Burley}, G.~S., \& {Kelson}, D.~D. 2014, {A Search for Stars
  of Very Low Metal Abundance. VI. Detailed Abundances of 313 Metal-poor
  Stars}.
\newblock {\em \aj}, 147(6), 136.

\bibitem[{Shetrone} et~al., 2003]{2003AJ....125..684S}
{Shetrone}, M., {Venn}, K.~A., {Tolstoy}, E., {Primas}, F., {Hill}, V., \&
  {Kaufer}, A. 2003, {VLT/UVES Abundances in Four Nearby Dwarf Spheroidal
  Galaxies. I. Nucleosynthesis and Abundance Ratios}.
\newblock {\em \aj}, 125(2), 684--706.

\bibitem[{Tang} et~al., 2020]{tang2020}
{Tang}, B., {Fern{\'a}ndez-Trincado}, J.~G., {Liu}, C., {Yu}, J., {Yan}, H.,
  {Gao}, Q., {Shi}, J., \& {Geisler}, D. 2020, {On the Chemical and Kinematic
  Consistency between N-rich Metal-poor Field Stars and Enriched Populations in
  Globular Clusters}.
\newblock {\em \apj}, 891(1), 28.

\bibitem[{Tang} et~al., 2019]{tang2019}
{Tang}, B., {Liu}, C., {Fern{\'a}ndez-Trincado}, J.~G., {Geisler}, D., {Shi},
  J., {Zamora}, O., {Worthey}, G., \& {Moreno}, E. 2019, {Chemical and
  Kinematic Analysis of CN-strong Metal-poor Field Stars in LAMOST DR3}.
\newblock {\em \apj}, 871(1), 58.

\bibitem[{Tang} et~al., 2023]{Tang2023}
{Tang}, B., {Zhang}, J., {Yan}, Z., {Zhang}, Z., {Carigi}, L., \&
  {Fern{\'a}ndez-Trincado}, J.~G. 2023, {Near-infrared chemical abundances of
  stars in the Sculptor dwarf galaxy}.
\newblock {\em \aap}, 669, A125.

\bibitem[{Tolstoy} et~al., 2009]{2009ARA&A..47..371T}
{Tolstoy}, E., {Hill}, V., \& {Tosi}, M. 2009, {Star-Formation Histories,
  Abundances, and Kinematics of Dwarf Galaxies in the Local Group}.
\newblock {\em \araa}, 47(1), 371--425.

\bibitem[{Xu} et~al., 2026]{Xu2026}
{Xu}, C., {Qiao}, Y., {Tang}, B., {Fern{\'a}ndez-Trincado}, J.~G., {Yan}, Z.,
  {Huang}, R., \& {Geisler}, D. 2026, {APOGEE chemical abundances of stars in
  the Milky Way satellites Fornax, Sextans, Draco, and Carina}.
\newblock {\em \aap}, 708, A259.

\bibitem[{Yan} et~al., 2017]{2017A&A...607A.126Y}
{Yan}, Z., {Jerabkova}, T., \& {Kroupa}, P. 2017, {The optimally sampled
  galaxy-wide stellar initial mass function. Observational tests and the
  publicly available GalIMF code}.
\newblock {\em \aap}, 607, A126.

\bibitem[{Yan} et~al., 2019]{2019A&A...629A..93Y}
{Yan}, Z., {Jerabkova}, T., {Kroupa}, P., \& {Vazdekis}, A. 2019, {Chemical
  evolution of elliptical galaxies with a variable IMF. A publicly available
  code}.
\newblock {\em \aap}, 629, A93.

\bibitem[{Yong} et~al., 2013]{2013ApJ...762...27Y}
{Yong}, D., {Norris}, J.~E., {Bessell}, M.~S., {Christlieb}, N., {Asplund}, M.,
  {Beers}, T.~C., {Barklem}, P.~S., {Frebel}, A., \& {Ryan}, S.~G. 2013, {The
  Most Metal-poor Stars. III. The Metallicity Distribution Function and
  Carbon-enhanced Metal-poor Fraction}.
\newblock {\em \apj}, 762(1), 27.

\end{thebibliography}

\end{document}